\documentclass[
 reprint,
 superscriptaddress,
prb,
showkeys,10pt
]{revtex4-2}
\usepackage{float}
\usepackage{multibib}
\usepackage{hyperref}
\usepackage{xcolor}
\hypersetup{
    colorlinks,
    linkcolor={blue!50!blue},
    citecolor={blue!50!blue},
    urlcolor={blue!80!black}
}
\usepackage{graphicx}
\usepackage{amsmath}
\usepackage[dvips]{epsfig}
\usepackage[sort&compress]{natbib}
\usepackage{physics}
\usepackage{color}
\usepackage{ulem}

\begin{document}

\title{Time-Resolved Stokes Analysis of Single Photon Emitters in Hexagonal Boron Nitride}  

\author{\c{C}a\u{g}lar Samaner}
\thanks{\href{mailto:caglarsamaner@iyte.edu.tr}{caglarsamaner@iyte.edu.tr} and \href{mailto:serkanates@iyte.edu.tr}{serkanates@iyte.edu.tr}.}
\affiliation{Department of Physics, \.{I}zmir Institute of Technology, 35430 \.{I}zmir, Turkey}
\author{Serkan Ate\c{s}}
\thanks{\href{mailto:caglarsamaner@iyte.edu.tr}{caglarsamaner@iyte.edu.tr} and \href{mailto:serkanates@iyte.edu.tr}{serkanates@iyte.edu.tr}.}
\affiliation{Department of Physics, \.{I}zmir Institute of Technology, 35430 \.{I}zmir, Turkey}
\affiliation{QLocked Technology Development Inc., 35430, İzmir, Turkey}
\begin{abstract}
Solid-state quantum emitters play a vital role in advancing quantum technologies, particularly in quantum computation and communication, where single-photon polarization acts as a fundamental information carrier. Precise polarization characterization is essential for understanding the mechanisms underlying polarization dynamics, which is critical for developing quantum emitters with minimized polarization-related errors. In this study, we employ the Rotating Quarter-Wave Plate (RQWP) method to comprehensively characterize the polarization state of quantum emitters in hexagonal boron nitride (hBN). By examining both time-averaged and dynamic polarization features, we demonstrate the time-resolved evolution of Stokes parameters from a solid-state single-photon emitter using the RQWP technique. This approach provides more complete polarization information than conventional micro-photoluminescence methods, without requiring modifications to the experimental setup. Our results uncover intricate polarization dynamics in hBN emitters, offering insights that were previously inaccessible. The techniques presented here can be broadly applied to polarization analysis of solid-state quantum emitters across various material platforms.
\end{abstract}


\maketitle

\section{Introduction}
Single-photon emitters (SPEs) are at the forefront of advancements in scalable quantum technologies due to their potential applications in on-chip photonic structures and circuits \cite{OBrien_2009, Lodahl2015, Uppu2021}. Various platforms, including semiconductor quantum dots \cite{Gallardo2010,Senellart2017}, nitrogen-vacancy centers in diamond \cite{Alegre2007,Bradac2019}, transition metal dichalcogenides \cite{He2015}, zinc oxide \cite{Morfa2012}, and silicon nitride \cite{Senichev2021}, are known to host SPEs. Hexagonal boron nitride (hBN) has become another promising material with its emergent \cite{Tran2016} and artificially created \cite{Kumar2023a} defect-based quantum emitters operable at room temperature. Despite the numerous platforms, the search for an efficient, reliable, and scalable single photon source remains a critical area of research due to its significant implications for practical applications \cite{Esmann2024}.

Single-photon nature and polarization of light emitted from SPEs are crucial in various quantum technologies, serving as fundamental components that utilize these unique properties for various applications \cite{Couteau_2023a, Couteau_2023b}. Notable examples include quantum metrology, where the non-classical properties of single photons enable enhanced precision in measurements for sensing and imaging techniques \cite{Muller_2017, Berchera2019}. In photonic quantum computing, SPEs are utilized as qubits, exploiting the principles of quantum superposition and entanglement to achieve computational capabilities that surpass classical systems \cite{ Maring_2024}. SPEs are also used in quantum random number generation applications utilizing either their quantum nature \cite{Chen2019} or other intrinsic random features such as symmetric dipole emission \cite{Hoese2022}. Moreover, in quantum key distribution (QKD), the polarization state of single photons facilitates the implementation of secure communication protocols, ensuring robust cryptographic security against eavesdropping \cite{Waks2002, Leifgen2014, Kupko2020, Samaner2022, Murtaza2023}. Most of the applications mentioned above utilize the polarization of single photons as a critical quantum feature to operate, thereby a thorough polarization characterization of such single-photon emitters becomes a crucial part of practical applications as well as fundamental research that uses such emitters. A complete polarization analysis may enhance the efficiency and reliability of quantum communication protocols by providing precise control over photon states. Additionally, understanding the polarization properties of SPEs may aid in unraveling the intricate behaviors of quantum systems. Moreover, polarization characterization can become a key factor for identifying defects in materials, optimizing SPE performance, and developing higher-quality quantum technologies.

In this work, we employ the Rotating Quarter Wave Plate (RQWP) method to analyze the polarization state of quantum emitters in hBN. Compared to conventional techniques that assess polarization only along linear bases, the RQWP method enables access to the full set of Stokes parameters, capturing both time-averaged and dynamic polarization features without compromising the simplicity of the experimental setup. This approach reveals previously inaccessible details of the polarization dynamics in hBN emitters~\cite{Kumar2024}, deepening our understanding of their fundamental optical properties and paving the way for improved performance in quantum photonic applications.

\begin{figure*}[!ht]
    \includegraphics[width=0.95\textwidth]{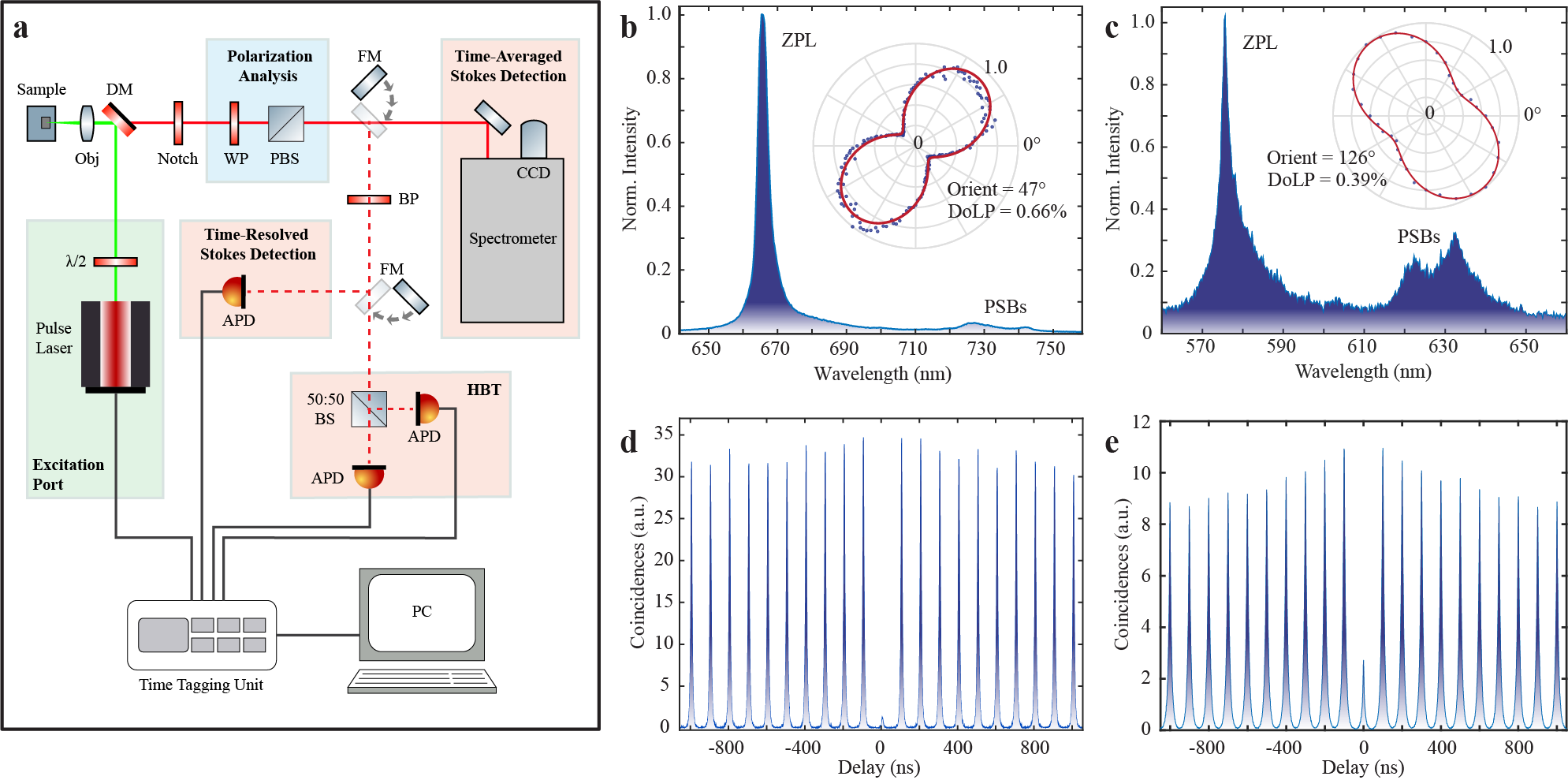}
    \caption{(a) Micro-photoluminescence setup that is used for the optical spectroscopy and polarization analysis of the emitters. (b) Normalized PL spectrum of emitter-1 at room temperature. The ZPL of the emitter is at 664~nm with characteristics PSB emission around 730~nm. The inset shows the linear polarization series of the ZPL taken by a HWP and PBS with a DoLP~=~0.66 and an orientation of linear polarization as $\Psi =47^\circ$. (c) Normalized PL spectrum of emitter-2 at room temperature. The ZPL of the emitter is at 575~nm with characteristics PSB emission around 630~nm. The inset shows the linear polarization series of the ZPL taken by a HWP and PBS with a DoLP~=~0.39 and an orientation of linear polarization as $\Psi =126^\circ$. (d) and (e) show present the result of second-order photon correlation measurement of ZPL emission from emitter-1 and emitter-2 with an antibunching value $g^{(2)}(0)~=~0.035~\pm0.002$ and $g^{(2)}(0)~=~0.36~\pm0.005$, respectively.}
    \label{fig-1}
\end{figure*}

\section{Results and Discussion}
Polarization of light is commonly described using two mathematical frameworks: the Jones and Stokes formalisms. For light propagating along a single axis (conventionally the \(z\)-axis), the electric field can be expressed as a complex two-component vector, known as the Jones vector. This formalism captures both amplitude and phase information and is applicable only to fully polarized light. In contrast, the Stokes formalism extends to partially polarized and unpolarized light by describing the polarization state in terms of time-averaged intensity components. The Stokes parameters \(S_0\), \(S_1\), \(S_2\), and \(S_3\) quantify the total intensity, linear polarization along horizontal/vertical, linear polarization at \(+45^\circ/-45^\circ\), and circular polarization (right versus left), respectively~\cite{Sharma_2006}.

Within this framework, the \textit{degree of polarization} (DoP) is defined as:
\begin{align}
\text{DoP} = \frac{\sqrt{S_1^2 + S_2^2 + S_3^2}}{S_0}
\label{eq:dop_main}
\end{align}
which represents the fraction of light that is polarized, satisfying \(0 \leq \text{DoP} \leq 1\). On the other hand, the \textit{degree of linear polarization} (DoLP), which is commonly used in experimental studies of single-photon sources, refers specifically to the linearly polarized component and is given by:
\begin{align}
\text{DoLP} = \frac{\sqrt{S_1^2 + S_2^2}}{S_0}
\label{eq:dolp_main}
\end{align}
indicating the projection of the DoP onto the linear polarization plane. This quantity satisfies the inequality \(0 \leq \text{DoLP} \leq \text{DoP} \leq 1\). If the light is purely linearly polarized, then \(\text{DoLP} = \text{DoP}\). However, in the presence of circular polarization components (\(S_3 \neq 0\)), \(\text{DoLP} < \text{DoP}\).  
Analyzing the polarization characteristics of a light source using the Stokes formalism offers a significant advantage over alternative polarization analysis techniques, which typically study only the linear projection of the polarization. The Stokes parameters provide all the information obtained through alternative methods, along with additional details, such as the ability to calculate the DoP for different projections (linear or circular) and to extract the completely polarized component from a partially polarized light beam.

\begin{figure*}[ht!]
   \includegraphics[width=0.95\textwidth]{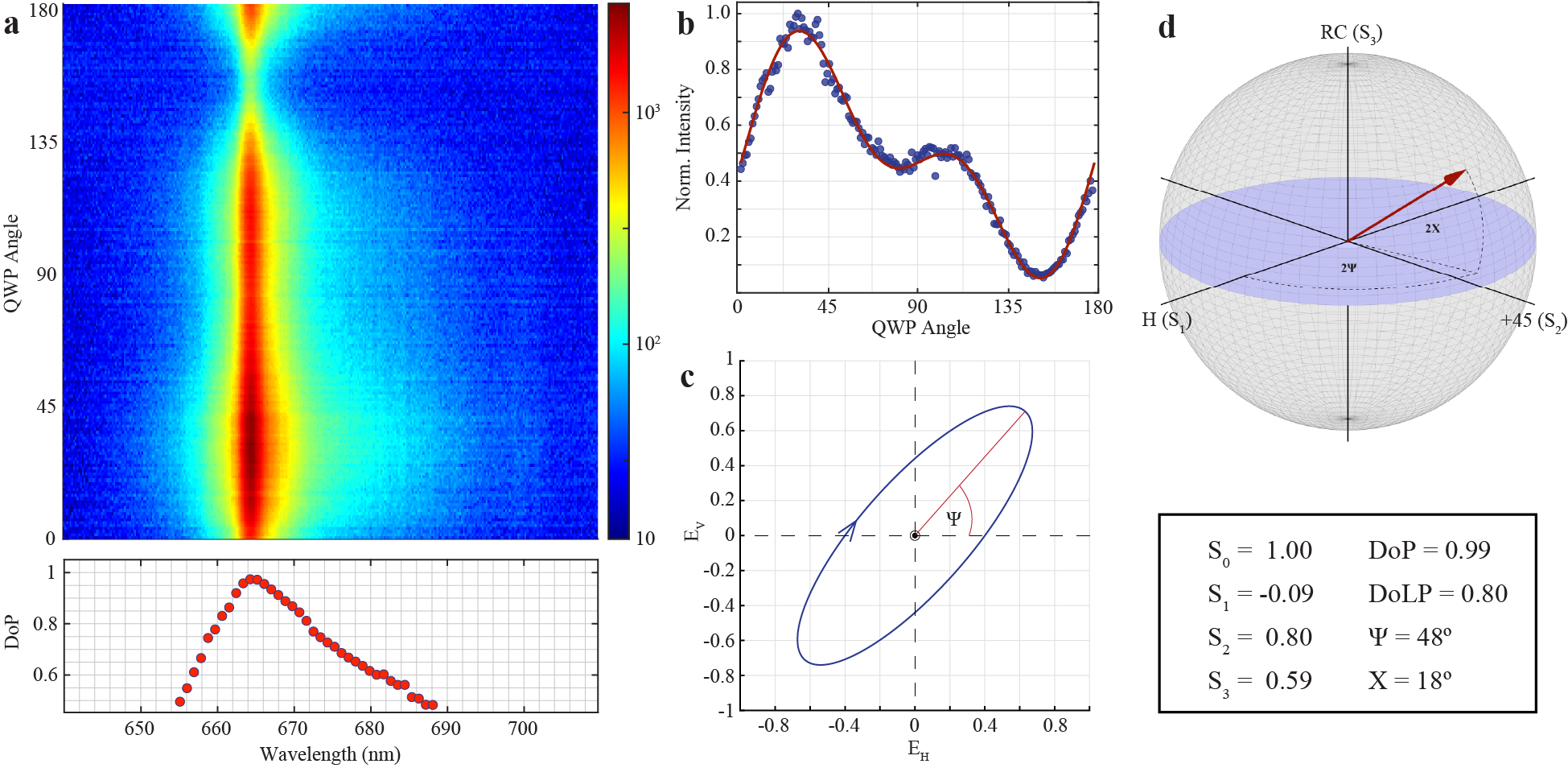}
    \caption{(a) Polarization map of emitter-1 performed using RQWP method. The spectrally resolved DoP of the emission around ZPL is shown at the bottom, with a resolution of 0.8 nm. (b) Integrated ZPL intensity as a function of QWP angle. Red solid curve shows the fit to the data to extract the Stokes parameters. (c) Polarization ellipse of the ZPL with $\Psi$ representing the orientation of the polarization with respect to horizontal. The blue arrow indicates the rotation of polarization due to circular contribution. (d) The polarization state of the ZPL shown on the Poincar\'e sphere where $X$ and $\Psi$ are the ellipse parameters. Axis labels H, +45, and RC indicate polarization bases of horizontal, diagonal, and right circular, respectively. Polarization-related parameters are displayed in the box at the bottom.}
    \label{fig-2}
\end{figure*}

Figure~\ref{fig-1}a illustrates the confocal micro-photoluminescence (\(\mu\)PL) setup used for optical spectroscopy, as well as for both time-averaged and time-resolved Stokes polarization analysis of the emitters. The setup also includes a Hanbury-Brown and Twiss (HBT) interferometer for photon correlation measurements. Details of the setup are provided in the Methods section. Here we present our experimental results using two different hBN emitters labeled emitter-1 and emitter-2. Figure~\ref{fig-1}b presents the room-temperature PL spectrum of emitter-1 under 483~nm pulsed excitation. The sharp peak at 664~nm is identified as the zero-phonon line (ZPL), while the smaller peaks around 730~nm correspond to two phonon sidebands arising from coupling to optical phonon modes of hBN~\cite{Cusco2016}. Linear polarization of the ZPL emission is measured using a combination of a half-wave plate (HWP) and a polarizing beam splitter (PBS) placed in the emission path, the result of which is shown in the inset. The data are fitted with a generic cosine-squared function to extract the dipole orientation \(\Psi~=~47^\circ\) and the DoLP~=~0.66. Figure~\ref{fig-1}c displays the room-temperature PL spectrum of emitter-2 under the same excitation conditions. A pronounced ZPL is observed at 575~nm, accompanied by phonon sidebands (PSBs) around 630~nm, indicative of coupling to optical phonon modes. The inset of Figure~\ref{fig-1}c presents the linear polarization analysis of the ZPL emission with dipole orientation of \(\Psi = 126^\circ\) and a DoLP = 0.39. To verify the single-photon nature of the emission, we performed second-order photon correlation measurements using the HBT interferometer for both emitters. The results are presented in Figures~\ref{fig-1}d and~\ref{fig-1}e for emitter-1 and emitter-2, respectively. The measured second-order correlation values at zero time delay are \(g^{(2)}(0) = 0.035 \pm 0.002\) for emitter-1 and \(g^{(2)}(0) = 0.36 \pm 0.005\) for emitter-2, clearly confirming strong photon antibunching and validating their operation as true single-photon sources.

\subsection{Time-averaged Stokes polarization analysis} 

\begin{figure*}[htp!]
    \includegraphics[width=0.95\textwidth]{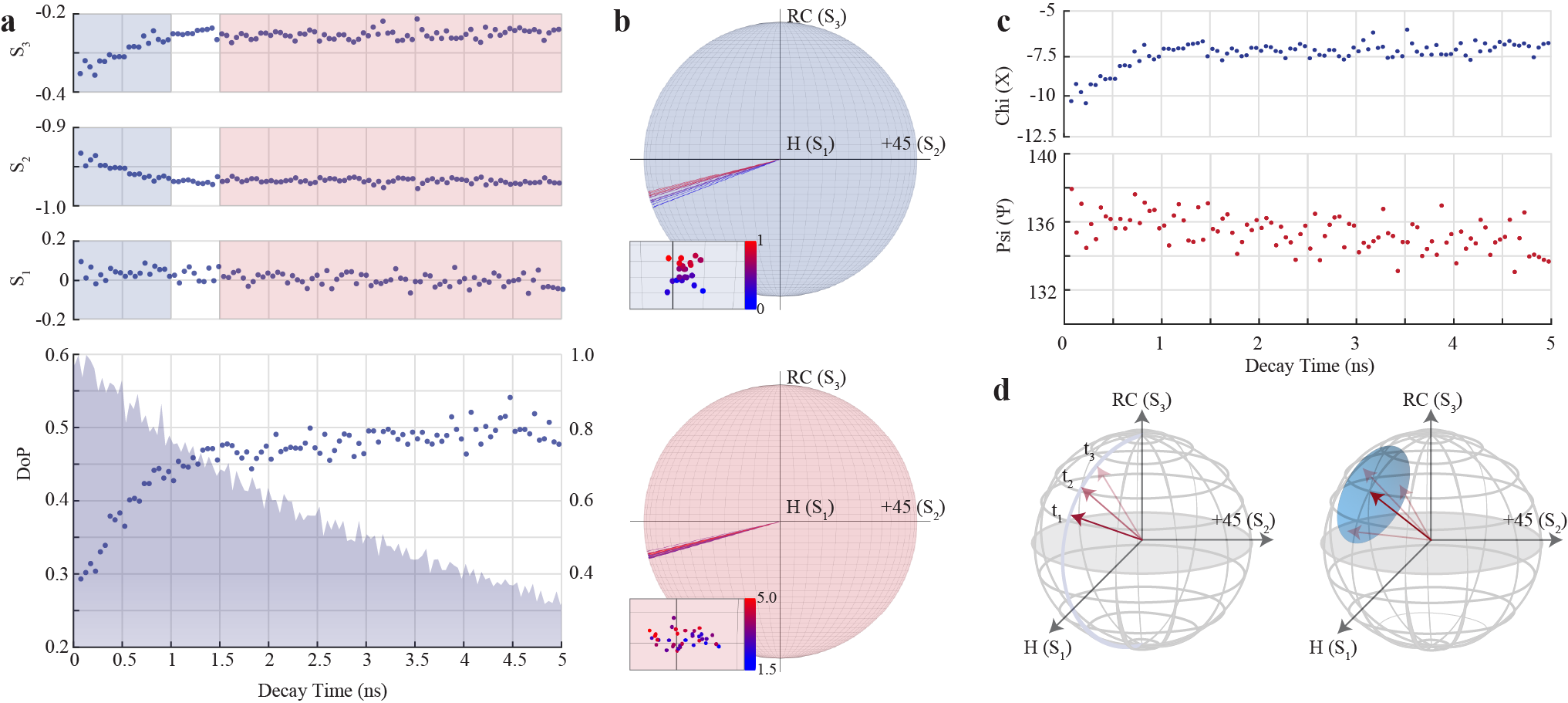}
    \caption{(a) Time-resolved DoP and Stokes parameters $S_1$, $S_2$, and $S_3$ of the completely polarized part of the emission, plotted with respect to the decay time of the emitter-2 for the first 5 ns after the excitation pulse. DoP plot also includes the result of time-correlated single photon counting measurement. (b) Polarization trace of emitted photons for two distinct regions considering the behavior of DoP. The region between 0 - 1 ns is colored blue, and the region between 1.5 - 5.0 ns is red. The arrows inside the spheres are also colored according to decay times. The inset shows the distribution of polarization states on the surface of the Poincar\'e sphere. (c) Dynamical behavior of ellipse parameters $\Psi$ and $X$ with respect to decay time of the emitter. (d) Two specific cases that could be responsible for the observed behavior of DoP. }
    \label{fig-3}
\end{figure*}

Here, we present the results of the Stokes polarization analysis for emitter-1, performed using the RQWP technique, which employs a quarter-wave plate (QWP) instead of a HWP before the PBS to extract the Stokes parameters and determine the DoP. Figure~\ref{fig-2} summarizes the polarization analysis conducted via the time-averaged measurement section of the experimental setup shown in Fig.~\ref{fig-1}. The ZPL polarization map, obtained by recording the PL spectrum of the emitter as a function of the QWP angle, is shown at the top of Fig.~\ref{fig-2}a. A validation of the Stokes analysis is provided in Fig.~\ref{fig-2}b, where the blue dots represent the integrated ZPL intensity at every QWP angle. The solid red line indicates the theoretical intensity curve calculated by applying the Mueller matrices of the QWP and PBS to the extracted Stokes parameters of the ZPL. The intensity of the light passing through these components can be written as~\cite{Schaefer2007}:

\begin{align}
I_n = \frac{1}{2} \left( A + B \sin 2\theta_n + C \cos 4 \theta_n + D \sin 4 \theta_n \right)
\label{intensityStokes}
\end{align}

\noindent where \(n\) is an integer ranging from 1 to \(N\) (\(N \geq 8\)), and \(\theta_n\) is the \(n\)th angle in the polarization measurement series, defined as the angle between the fast axis of the QWP and the horizontal axis. Here, coefficients A, B, C, and D are related to the Stokes parameters (see Methods).

The polarization ellipse of the ZPL is depicted in Fig.~\ref{fig-2}c. Here, the symbol $\Psi$ denotes the orientation of the polarization axis relative to the horizontal, while the arrow illustrates the direction of rotation induced by the circular polarization component. Finally, the polarization state of the ZPL is visualized on the Poincaré sphere in Fig.~\ref{fig-2}d, with the extracted Stokes parameters listed in the box below: $S_0 = 1.00$, $S_1 = -0.09$, $S_2 = 0.80$, and $S_3 = 0.59$. The corresponding values of the degree of polarization, $\mathrm{DoP} = 0.99$, and the degree of linear polarization, $\mathrm{DoLP} = 0.80$, calculated using Eqs.~\ref{eq:dop_main} and~\ref{eq:dolp_main}, indicate a significant difference due to the presence of a circular polarization component. The parameters of the polarization ellipse are determined to be $\Psi = 48^\circ$ and $X = 18^\circ$, where $\Psi$ represents the polarization orientation and $X$ denotes the ellipticity. A comparison of $\Psi$ with the value given in the inset of Fig.~\ref{fig-1}b confirms excellent agreement between the two independent measurements. 

The results show that although the emitted light is nearly completely polarized, the measured DoLP is 80\% due to a non-negligible circular polarization component. Such reduced linear polarization visibility is commonly observed in single-photon emitters in hBN~\cite{Tran2016,Patel2022,Ziegler2018}. The bottom panel of Fig.\ref{fig-2}a presents the spectrally resolved DoP, extracted by binning the wavelength axis in the polarization map and performing Stokes analysis. The ZPL exhibits a high DoP of up to 99\% at its spectral center (664nm), indicating that emission is completely polarized. However, the DoP decreases for wavelengths detuned from the ZPL, where phonon-assisted processes dominate. These involve interactions with optical and acoustic phonons, which introduce decoherence and couple to vibrational modes of varying symmetry, thus reducing polarization purity~\cite{Exarhos2017, Jungwirth2016}. This behavior is consistent with previous observations of phonon-induced depolarization in hBN emitters~\cite{Jungwirth2016a}.

\subsection{Time-resolved Stokes polarization analysis}
In addition to the time-averaged Stokes analysis performed on emitter-1, we conducted, for the first time, a time-resolved Stokes analysis on a single defect in hBN using the RQWP technique to investigate the dynamic evolution of polarization. This method involves recording photoluminescence decay curves at multiple QWP angles and extracting the Stokes parameters as a function of time after excitation. Using emitter-2, Figure~\ref{fig-3}a presents the degree of polarization (DoP) and the corresponding Stokes parameters of the fully polarized emission component—isolated after filtering out any unpolarized or partially polarized contributions—over a 5 ns time window. Notably, the DoP increases rapidly within the first nanosecond and then stabilizes, consistent with previous observations for defects in hBN and diamond~\cite{Kumar2024}. During this initial period, the Stokes parameters $S_2$ and $S_3$ undergo significant changes, while $S_1$ fluctuates around zero with relatively higher variation.

Figure~\ref{fig-3}b shows the polarization states as vectors inside the Poincaré sphere, corresponding to early (blue) and late (red) emission time windows. These states are illustrated on the sphere’s surface with a color gradient representing relative decay time. During the first nanosecond, the polarization state evolves toward the linear polarization plane, as evidenced by decreasing circular components ($S_3$) and evolving linear diagonals ($S_2$). In the later time window (red), $S_1$ fluctuations persist, while $S_2$ and $S_3$ remain relatively stable. Figure~\ref{fig-3}c shows the evolution of the polarization ellipse parameters $\Psi$ and $X$, derived from the Stokes parameters, reflecting changes in the ellipticity and orientation of the emitted light over time.

Based on these observations, we identify two underlying mechanisms, illustrated in Fig.\ref{fig-3}d. On the left, the early-time polarization dynamics suggest a relaxation process occurring within the first 1.5~ns post-excitation. On the right, persistent fluctuations in the polarization state are observed throughout the emission period, with greater variation in the linear polarization plane. These effects may arise from environmental factors such as surface charges or spectral diffusion~\cite{Exarhos2017}.

The increase in DoP from approximately 0.3 to 0.5 during the initial 1.5~ns (Fig.\ref{fig-3}a) reflects a temporal evolution from a partially mixed polarization state toward a more coherent emission. This evolution likely originates from relaxation among excited sublevels or phonon-coupled states. Immediately after excitation, the emitter may occupy a superposition of excited states with differing dipole orientations. As the system relaxes into a lower-energy excited state with a better-defined dipole, polarization coherence improves, resulting in a higher DoP. Similar behavior has been reported in hBN defects and NV centers in diamond~\cite{Kumar2024}. However, the saturation of DoP at 0.5, rather than reaching unity, points to a persistent depolarization mechanism. Possible causes include fast spectral diffusion, phonon-assisted transitions, or emission from unresolved fine-structure levels. Additionally, fluctuations in the local environment—such as surface fields or strain—may further contribute to depolarization during the emission process~\cite{Jungwirth2016}. These findings underscore the importance of time-resolved polarization measurements in revealing the intrinsic dynamics of solid-state quantum emitters. Compared to conventional time-averaged polarization measurements, the RQWP-based time-resolved Stokes analysis offers a more comprehensive view of the emission polarization, enabling simultaneous access to the full Stokes parameters and their dynamic evolution. This approach is particularly powerful in revealing polarization relaxation mechanisms and environmental effects that govern the optical behavior of quantum emitters in solid-state hosts such as hBN.

\section{Conclusions}

We employed the RQWP method to perform a full Stokes polarization analysis of single-photon emitters in hBN. This technique enabled a detailed characterization of both time-averaged and time-resolved polarization dynamics, revealing features not captured by conventional linear polarization measurements. Our time-averaged analysis shows a circular polarization component, often neglected in the literature, which can aid in emitter identification schemes that correlate polarization with crystal axes \cite{Zhong2024,horder2024nearcoherent}. Our results also reveal that less than unity DoLP may originate simply from the polarization state of the emitter, and the true DoP should be considered when investigating polarization-related properties of emitters alongside other responsible mechanisms \cite{Jungwirth2017,Wang2017,Rugar2019}. Time-resolved measurements further reveal that the emitter’s polarization traces a distinct trajectory on the Poincar\'e sphere during the first nanosecond of decay, driven by a decrease in circular polarization. This behavior offers valuable insight into emitter dynamics and may help to optimize polarization properties for quantum applications. These results highlight the strength of the RQWP method as a simple yet powerful tool for complete polarization characterization. Its compatibility with time-resolved studies makes it especially valuable for advancing solid-state quantum emitters in quantum communication technologies.

\section*{Methods}
\subsection*{Sample preparation}
Multilayer hBN flakes, commercially obtained from Graphene Supermarket as a solution with 5.5 mg/L concentration. Approximately 10 $\mu$L of this solution was drop-cast onto a pre-cleaned Si/SiO$_2$ substrate. The sample was then allowed to dry naturally under ambient conditions, facilitating the gradual evaporation of the solvent and the adherence of the hBN flakes to the substrate.

\subsection*{Experimental Setup} 
The optical characteristics of hBN quantum emitters were investigated using a custom-built confocal micro-photoluminescence setup, as depicted in Figure \ref{fig-1}a. Optical measurements were conducted with a 483~nm pulsed laser (Pilas, Advanced Laser Diode Systems) featuring a pulse duration of less than 80~ps (FWHM) in the excitation port. An HWP positioned after the pulsed laser was used to maximize the excitation efficiency of the emitters. Excitation and emission collection were achieved using an objective lens with a numerical aperture of 0.90 and 100x magnification (M Plan Apo HR, Mitutoyo). A notch filter placed after the dichroic mirror ensured the suppression of residual laser light. All polarization measurements were carried out using an RQWP analyzer, which consisted of a combination of a QWP and a PBS, except for the linear polarization analysis presented in Figure \ref{fig-1}c, where an HWP was used in place of the QWP. The detection port was separated into three sections, highlighted in light-red in Figure \ref{fig-1}a, where the light was directed using a series of flip mirrors. Time-averaged Stokes polarization analysis was performed by combining a spectrometer with a resolution of 0.03~nm (Shamrock 750, Andor) and a charge-coupled device (CCD) camera (Newton, Andor). Measurements were taken with a 1-second exposure time for the CCD camera at each QWP angle. To spectrally isolate the emitter signal before conducting time-resolved or time-correlated measurements, a commercially available bandpass filter was employed. Time-resolved Stokes polarization measurements were conducted using a single avalanche photodiode (SPCM-AQRH, Excelitas), with 1-minute integration times at each QWP angle. A time-tagger unit (quTAQ HR, Qutools) was used to record the excitation and detection times during the time-resolved polarization measurements. After the measurements, data was post-processed to calculate the time-resolved Stokes parameters. Finally, a Hanbury Brown and Twiss interferometer was utilized to perform second-order photon correlation experiments.

\subsection{Stokes Polarization Analysis}
\subsubsection{Stokes Polarization Parameters}
Two widely adopted mathematical frameworks are used to represent and study the polarization of light. In most experimental cases, light propagates in a single direction, conventionally chosen as the \(z\)-axis. Such electromagnetic radiation has no electric field components along the \(z\)-axis, allowing it to be represented by two-component vectors known as Jones vectors. These vectors have complex-valued components, making them essential for analyzing interference effects. However, a key limitation of the Jones formalism is that it applies only to completely polarized light, where the amplitude and phase of the electric field remain constant over time.

To describe partially polarized or unpolarized light, an alternative representation, known as the Stokes formalism, must be used. In this framework, quasi-monochromatic light is expressed in terms of time-averaged field intensities using Stokes parameters as \cite{Sharma_2006}:

\begin{align}
S = \begin{bmatrix} S_0 \\ S_1 \\ S_2 \\ S_3 \end{bmatrix} = 
\begin{bmatrix} 
\langle \epsilon_{x}^{2}(t) \rangle + \langle \epsilon_{y}^{2}(t) \rangle \\ 
\langle \epsilon_{x}^{2}(t) \rangle - \langle \epsilon_{y}^{2}(t) \rangle \\ 
2 \langle \epsilon_x (t) \epsilon_y (t) \cos(\phi (t)) \rangle \\ 
2 \langle \epsilon_x (t) \epsilon_y (t) \sin(-\phi (t)) \rangle  
\end{bmatrix}
\label{StokesPars}
\end{align}

Equation \ref{StokesPars} represents the real-valued field amplitudes in the \(x\)- and \(y\)-directions, with \(\phi\) denoting the relative phase difference between these components. The Stokes parameters \(S_0\), \(S_1\), \(S_2\), and \(S_3\) quantify different aspects of the polarization of the light. Specifically, \(S_0\) corresponds to the total irradiance of light, while \(S_1\) measures the excess of horizontal over vertical polarization. The parameter \(S_2\) quantifies the excess of \(+45^\circ\) over \(+135^\circ\) polarization, and \(S_3\) represents the difference between the circular right polarization (RCP) and circular left polarization (LCP). It is important to note that the Stokes parameters satisfy the following fundamental inequality.

\begin{align}
S_{0}^2 \geq S_{1}^2 + S_{2}^2 + S_{3}^2 \label{StokesParsQuasiMonoChromatic}
\end{align}

\noindent where equality holds for completely polarized light, in which the total irradiance is equal to the sum of squares of $S_1$, $S_2$, and $S_3$. In contrast, the inequality indicates partially polarized light, where the sum of the squares of $S_1$, $S_2$, and $S_3$ is less than the total irradiance $S_{0}^2$. The DoP of a light beam is then defined in terms of the Stokes parameters as follows:

\begin{align}
    DoP = \frac{\sqrt{S_{1}^2 + S_{2}^2 + S_{3}^2}}{S_0}~with~\ 0 \leq DoP \leq 1 \label{DoP}
\end{align}

According to Equation~\ref{DoP}, light is classified as completely polarized when $DoP = 1$ and unpolarized when $DoP = 0$. Any value between 0 and 1 represents partially polarized light. Similarly, the DoLP can be defined in the same way as the DoP, as follows:

\begin{align}
DoLP = \frac{\sqrt{S_{1}^2 + S_{2}^2}}{S_0} = \frac{I_{Max} - I_{Min}}{I_{Max} + I_{Min}}
\label{DoLP}
\end{align}

Here, $I_{\text{Max}}$ and $I_{\text{Min}}$ represent the maximum and minimum intensities observed after the light beam passes through a rotating polarizer \cite{Chipman_Lam_Young_2018}. The DoLP is defined as the projection of the DoP onto the linear plane. Consequently, it satisfies the following relation:

\begin{align}
0 \leq DoLP \leq DoP \leq 1
\label{DoLPineq}
\end{align}

If the polarization of light is linear, the DoLP will be equal to the DoP. However, if any circular polarization components are present, the DoLP will be less than the DoP.

Stokes analysis also allows for the decomposition of partially polarized light into its completely polarized and unpolarized components:

\begin{align}
\begin{bmatrix} S_0\\ S_1\\ S_2\\ S_3 \end{bmatrix} = (1-DoP) \begin{bmatrix} S_0\\ 0\\ 0\\ 0 \end{bmatrix} + \begin{bmatrix} DoP*S_{0} \\ S_1\\ S_2\\ S_3 \end{bmatrix} \label{decomp}
\end{align}

The first term on the right-hand side represents the unpolarized component, where $S_1$, $S_2$, and $S_3$ are all zero, while the second term corresponds to the completely polarized part. This decomposition allows us to analyze the polarization characteristics of the fully polarized portion of the light beam.

\subsubsection{Rotating Quarter Wave-plate Method}
In this subsection, we briefly introduce the RQWP method for measuring the Stokes parameters~\cite{Schaefer2007}. This method uses a QWP and a polarizing beam splitter to characterize the polarization of light. The intensity of the light passing through these components can be written as:

\begin{align}
I_n = \frac{1}{2} \left( A + B \sin 2\theta_n + C \cos 4 \theta_n + D \sin 4 \theta_n \right)
\label{intensityStokes}
\end{align}

\noindent where \(n\) is an integer ranging from 1 to \(N\) (\(N \geq 8\)), and \(\theta_n\) is the \(n\)th angle in the polarization measurement series, defined as the angle between the fast axis of the QWP and the horizontal axis. Here, coefficients A, B, C, and D are related to the Stokes parameters by the following relations:

\begin{align}
S_0 &= A - C, \ \ \ \ \ S_1 = 2C,\ \ \ \ \ \\ S_2 &= 2D, \ \ \ \ \ \ \ \ \ S_3 = B. \nonumber \label{aBCD2Stokes}
\end{align}

Coefficients A, B, C, and D can be determined by the following relations:

\begin{align}
A &= \frac{2}{N} \sum_{n=1}^{N} I_n,\ \ &B=\frac{4}{N}\sum_{n=1}^{N} I_n\ sin\ 2\theta_n,  \\ 
C &= \frac{4}{N} \sum_{n=1}^{N} I_n\ cos\ 4\theta_n,\ \ &D=\frac{4}{N}\sum_{n=1}^{N} I_n\ sin\ 4\theta_n, \nonumber \label{aBCD} 
\end{align}

where N is the total number of measurements in the series, bigger than 8 and an even number.

Analyzing the polarization characteristics of a light source using the Stokes formalism with the RQWP offers a significant advantage over alternative polarization analysis techniques, which typically study only the linear projection of the polarization. This advantage arises from the fact that no trade-offs are required in the experimental setup; the only modification is replacing the HWP with a QWP. The Stokes parameters provide all the information obtained through alternative methods, along with additional details, such as the ability to calculate the DoP for different projections (linear or circular) and to extract the completely polarized component from a partially polarized light beam.





\begin{acknowledgments}
This work was supported by the QuantERA II Programme that has received funding from the EU Horizon 2020 research and innovation programme under GA No 101017733 (Comphort), Scientific and Technological Research Council of Turkey (TUBITAK) under GA Nos. 124N115 and 124N110 and Izmir Institute of Technology under Grant No. 2023IYTE-2-0017. S.A. acknowledges the support from the Turkish Academy of Sciences (TUBA-GEBIP) and the BAGEP Award of the Science Academy.
\end{acknowledgments}

\bibliography{main}

\end{document}